\documentclass[aps,prb, twocolumn,floatfix,notitlepage, superscriptaddress,longbibliography]{revtex4-2}
\usepackage[utf8]{inputenc}
\usepackage{amsfonts,amsmath,amssymb,graphicx,epstopdf,verbatim}
\usepackage{rotating}
\usepackage{epstopdf}
\usepackage{xcolor}
\usepackage{natbib}
\usepackage[colorlinks]{hyperref}
\definecolor{darkred}{rgb}{0.5,0,0}
\definecolor{darkgreen}{rgb}{0,0.5,0}
\definecolor{darkblue}{rgb}{0,0,0.5}
\hypersetup{colorlinks,
linkcolor=darkred,
filecolor=darkgreen,
urlcolor=darkblue,
citecolor=darkgreen}
\usepackage{braket}

\newcommand{\tr}{\mathrm{Tr}}

\newcommand{\calN}{\mathcal{N}}

\newcommand{\ud}{\mathrm{d}}
\newcommand{\nep}{\operatorname{e}}

\newcommand{\opcdag}[1]{{\hat{c}^{\dagger}}_{#1}}

\newcommand{\opc}[1]{{\hat{c}^{\phantom \dagger}}_{#1}}

\begin{document}

\title{Entanglement transitions in the quantum Ising chain: \\
  A comparison between different unravelings of the same Lindbladian}

\author{Giulia Piccitto}

\affiliation{Dipartimento di Fisica dell’Università di Pisa and INFN, 
Largo Pontecorvo 3, I-56127 Pisa, Italy}

\author{Angelo Russomanno}
\affiliation{Scuola Superiore Meridionale, Università di Napoli Federico II
Largo San Marcellino 10, I-80138 Napoli, Italy}

\author{Davide Rossini}
\affiliation{Dipartimento di Fisica dell’Università di Pisa and INFN, 
Largo Pontecorvo 3, I-56127 Pisa, Italy}

\begin{abstract}
  We study the dynamics of entanglement in the quantum Ising chain with dephasing dissipation
  in a Lindblad master equation form. We consider two unravelings which preserve the Gaussian form of the state,
  allowing to address large system sizes. The first unraveling gives rise to a quantum-state--iffusion dynamics,
  while the second one describes a specific form of quantum-jump evolution, suitably constructed to preserve Gaussianity.
  In the first case we find a finite-size crossover from area-law to logarithm-law entanglement scaling when varying the external field strength and the measurement rate, and draw the related phase diagram.
  In the second case we find that this crossover moves to parameters different than before, thus modifying the phase diagram. This evidences a different entanglement behavior
  for different unravelings of the same Lindblad equation. Finally, we compare these outcomes with the predictions
  of a non-Hermitian Hamiltonian evolution, finding conflicting results.
\end{abstract}

\maketitle

\section{Introduction}
\label{sec:intro}
Understanding the physics of quantum systems coupled to an external environment is intriguing both for applications
in recent quantum technologies and from a fundamental perspective, being related to the boundary
between classical and quantum domains~\cite{feynman,Leggett,Zurek}. 
In the hypothesis of a weak and Markovian coupling with the bath, the underlying dynamics can be
reliably described by a master equation in the Lindblad form~\cite{Manzano2020, Morigi2002, Petruccione}
for the reduced density matrix of the system. 
Recently it has been argued that this framework can also model the process of quantum measuring at random or continuous times. 
In fact, the external environment can be thought as a measurement apparatus performing measurements on the quantum system~\cite{Daley2014,Plenio}. 
While the measurement process itself is stochastic and the outcome is a pure-state quantum trajectory,
the density matrix obtained by averaging over such trajectories obeys a Lindblad-type evolution.
The evidence that different measuring protocols commonly lead to different behaviors~\cite{Jenograk2021} have renewed the interest in the measurement problem.
 
Here we focus on entanglement, a crucial property for quantum computing purposes~\cite{Nielsen, rossini} with a key role in quantum many-body systems~\cite{Amico_RMP}, and study the effect on the entanglement dynamics of the coupling to a classical measurement apparatus performing random measurements. 
In particular, we study the so-called measurement-induced entanglement transitions:
When measuring a quantum many-body system, the interplay between the unitary dynamics (contributing to the creation
of entanglement) and the quantum measurements (contributing to its destruction)
might result in sharp transitions between different dynamical phases, characterized by qualitatively different
entanglement properties in the asymptotic regime~\cite{DeLuca2019, Skinner2019, Chan2019, Szynieszewski2019, Fuji2020, Goto2020, Lunt2020, Nahum2020, Lang2020, Alberton2021, Botzung2021, Collura2021}.

A paradigmatic example is provided by random circuits undergoing random measurements~\cite{Li2018, Li2019, Gullans2020, Jian2020, Zabalo2020, Choi2020, Bao2020, Zhang2020, Romito2020, Turkeshi2020, Li2021, Li2021_2, Fan2021, Lunt2021, Lavasani2021, Sang2021, Shtanko2020, Lavasani2020, Shi2020, Block2021, Vasseur2021},
which may display a transition between a phase with volume-law scaling of entanglement and another phase with area-law scaling.
In the case of free-fermion Hamiltonians, the asymptotic volume-law of the unitary evolution~\cite{Fagotti2008,Calabrese2005,Vidal2003} is unstable for any measurement rate and exhibits a transition towards a subextensive phase~\cite{DeLuca2019,Skinner2019,Alberton2021,Collura2021}. This transition is observed even for long-range free-fermion Hamiltonians~\cite{Block2021}. 
Similar results have been obtained for free-fermion random circuits with temporal randomness~\cite{Chen2020}, a setting that has been recently generalized to higher dimensions~\cite{Tang2020}, Majorana random circuits~\cite{Bao2021,Nahum2020}, and Dirac fermions~\cite{Buchhold2021}. 
These entanglement transitions can sometimes be predicted by a related non-Hermitian Hamiltonian evolution, as discussed in Ref.~\cite{Turkeshi2021} for a quantum Ising chain with no transverse field. 

The set of stochastic trajectories whose average gives rise to a Lindblad evolution (unraveling) depends on the physics of the involved measurement process~\cite{Plenio}.
A question which has been only marginally addressed so far is related to the observation that any Lindblad equation
has many possible unravelings. The entanglement properties are encoded in the stochastic quantum trajectories, that,
contrarily to the average density matrix, contain only quantum correlations. One expects these correlations
to strongly depend on the specific unraveling.
In fact, each unraveling corresponds to a specific measurement process and a specific way
for destroying the entanglement generated by the unitary part of the dynamics.
This can give rise to different dynamical phases and entanglement transitions.
An example of that is provided in Ref.~\cite{DeLuca2019}, where the unitary unraveling gives rise to
a volume-law asymptotic entanglement, while the quantum-state-diffusion one provides an area-law entanglement.

\begin{figure}[!t]
  \includegraphics[width=0.45\textwidth]{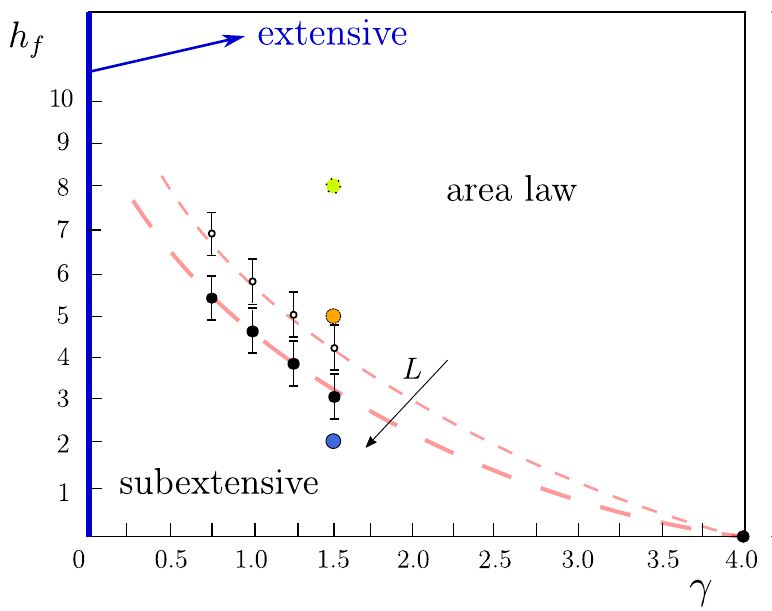}
  \caption{Sketchy phase diagram of the measurement-induced entanglement transition in the quantum Ising chain,
    obtained by solving the quantum-state-diffusion dynamics. Here $\gamma$ denotes the measurement rate,
    while $h_f$ the transverse-field strength. The line $\gamma = 0$ corresponds to the singular limit
    of the unitary dynamics, where the asymptotic entanglement entropy obeys a volume-law scaling (i.e., it
    grows proportionally to the system size $L$).
    For $\gamma > 0$, the entanglement is either subextensive (i.e., it scales $\propto \log L$) or it
    follows an area-law scaling (i.e., it is independent of $L$).
    The crossover between these two regimes occurs at a threshold value of $h_f$
    which strongly depends on the system size and on the measurement rate.
    Black dots with error bars denote the threshold fields evaluated at $L_\text{max} = 256$ (big filled circles) and $L_\text{max} = 192$ (small empty circles).
    Red dashed lines are guides to the eye. 
    The blue, orange, and light-green circles mark the parameters used for simulating the quantum trajectories
    dynamics in Fig.~\ref{Fig:ee_qj}. With this unraveling, the entanglement entropy exhibits an area-law phase even for parameter chosen deep in the subextensive region} 
  \label{Fig:tdf}
\end{figure}

In this paper we study measurement-induced quantum transitions in the transverse-field quantum Ising chain
undergoing a dephasing measurement process.
We quantify entanglement using mainly the entanglement entropy~\cite{RevModPhys.81.865}, a well-known entanglement monotone, and consider two different unravelings: First we solve the quantum-state-diffusion stochastic Schr\"odinger equation,
then we consider an unraveling based on a quantum-jump measurement process.
The phase diagram in Fig.~\ref{Fig:tdf}, obtained by solving the quantum-state-diffusion dynamics,
extends the results of Ref.~\cite{Turkeshi2021} to the case of non-zero transverse field. 
We find the existence of a crossover point from a subextensive to an area-law phase that depends
on the measurement and the field strength; the subextensive region reduces while increasing the system size.
For small couplings with the measurement apparatus, we observe the emergence of a maximum
in the asymptotic entanglement entropy close to the zero-temperature quantum critical point,
as a reminiscence of the unitary dynamics.

We then compare these results with those obtained from the quantum-jump measurement process. 
To this purpose, we choose slightly different measurement operators, which preserve the Gaussian form
of the state (allowing numerics for quite large system size) and at the same time result in the same Lindblad equation. 
With this unraveling we find that the entanglement entropy transition moves to parameters different than those predicted by the quantum-state-diffusion.
In fact, we monitor the entanglement trajectories at fixed $\gamma = 1.5$, for $h_f = 2,\  5,\  8$,
marked in Fig.~\ref{Fig:tdf} by the blue, orange, and light-green circle, respectively,
and always observe an area-law behavior.
However we should stress that, considered the size-dependence of the crossover point and the computational effort
needed for simulations, it is not possible to completely rule out the existence of an area-law phase
from a numerical analysis.
Finally, we compare both unravelings with the non-Hermitian Hamiltonian dynamics 
and find no agreement with this approximate dynamics.

The paper is organized as follows. In Sec.~\ref{sec:tf} we introduce the model, 
the two unravelings, and the way they preserve the Gaussianity of the state in each quantum trajectory. We also introduce
the entanglement entropy and discuss how it can quantify the entanglement for different unravelings.
In Sec.~\ref{Sec:Rqsd} we discuss our results on the quantum-state-diffusion unraveling,
in particular its phase diagram. 
In Sec.~\ref{Sec:Rqj} we present the asymptotic entanglement in the quantum-jump unraveling.
We also show that results with both unravelings cannot be predicted through a non-Hermitian Hamiltonian evolution.
Finally, in Sec.~\ref{sec:conclusions} we draw our conclusions.
Appendix~\ref{entone:sec} contains the results for the second R\'enyi entropy and the square equal-time correlation function. They are other two nonlinear functions of the projector on the state, and witness the entanglement transition in the quantum-state-diffusion unraveling.
The other Appendixes contains technical details that are useful for the numerical treatment of the
dynamics of Gaussian fermionic systems in the presence of quantum measurements.
%
\section{Theoretical framework}
\label{sec:tf}

\subsection{Hamiltonian model}
\label{subsec:Hamiltonian}

We consider the integrable quantum Ising chain with $L$ spins, described by the Hamiltonian
\begin{equation}
  \hat H = - J\sum_{j=1}^L \hat \sigma_j^x \hat \sigma_{j+1}^x - h \sum_{j=1}^L \hat \sigma_j^z.
	\label{Eq:ising}
\end{equation}
The spin-$1/2$ Pauli matrices $\hat \sigma_j^\alpha$ act on the $j$th site ($\alpha = x,y,z$),
while $J$ and $h$ denote, respectively, the spin-spin coupling and the transverse magnetic-field strength.
In what follows we set $J = 1$ as the energy scale, assume 
periodic boundary conditions ($\hat \sigma_{L+1}^\alpha \equiv \hat \sigma_1^\alpha$),
and work in units of $\hbar = 1$.
The Hamiltonian~\eqref{Eq:ising} features a parity symmetry generated by the operator
$\hat P = \otimes_{j = 1}^L \hat \sigma_j^z$, which divides the Hilbert space into two subspaces
of dimension $2^{L-1}$. 
The quantum Ising chain is known to exhibit a zero-temperature quantum phase transition
from a paramagnetic state (for $h > h_c$) to a ferromagnetic state (for $h < h_c$), when the transverse-field
strength crosses the critical value $h_c = 1$ and the above symmetry is broken~\cite{Sachdev2011}.
Without loss of generality we restrict our analysis of the dynamics to the even parity sector.

The ground-state entanglement entropy (see Sec.~\ref{ent:sec}) of a subchain with $\ell$ sites
obeys an {\em area-law} scaling (meaning that it is independent of $\ell$, in one-dimension),
except at the critical point, where it grows logarithmically with $\ell$~\cite{Calabrese2004,Vidal2003}.
On the other hand, during the unitary dynamics after a sudden quench in the transverse field $h$ (or during a periodic driving~\cite{russomanno2016}),
it exhibits a linear growth in time eventually attaining an asymptotic constant value which increases
linearly with $\ell$ (i.e., {\em volume-law} behavior, in one dimension)~\cite{Calabrese2005}.
The asymptotic entanglement entropy as a function of $h$ features a non-analytical cusp at $h_c$. 

As explained in Appendix~\ref{App:ising}, the above Ising chain~\eqref{Eq:ising}
can be mapped into a quadratic spinless-fermion Hamiltonian.
Using the Nambu spinor notation,
$\hat \Psi = (\hat c_1, \cdots, \hat c_L, \hat c_1^\dagger, \cdots, \hat c_L^\dagger)^T$,
with $\hat c_j^{(\dagger)}$ denoting anticommuting fermionic annihilation (creation) operators,
such Hamiltonian reads
\begin{equation}
  \hat H = \tfrac{1}{2} \hat \Psi^\dagger \mathbb{H} \hat \Psi + \text{const.}\,,
  \label{Eq:isingFermi}
\end{equation}
where $\mathbb{H}$ is the so-called Bogoliubov-De Gennes matrix defined in Eq.~\eqref{Eq:bdg}. 
This can be easily diagonalized by performing a $2L \times 2L$ transformation $\mathbb{U}$ such that
\begin{equation}
  \mathbb{U}^{-1} \mathbb{H} \mathbb{U} = \text{diag}(\omega_k, -\omega_k)\,. 
  \label{eq:UVdiag}
\end{equation}
For the Hamiltonian in Eq.~\eqref{Eq:ising}, the dispersion relation is
\begin{equation}
  \omega_k = 2 J \sqrt{1 + (h/J)^2 -2(h/J) \cos k},  
  \label{Eq:eigv}
\end{equation}
where the momenta $k = 2\pi n/L, \ n = -L/2 + 1, \dots, L/2$ are fixed by the parity sector we chose.
This analysis will be useful in Sec.~\ref{Subsec:noclick}, in the context of the non-Hermitian Hamiltonian approximation.

A similarly simple treatment is possible for a non-uniform model and for the nonequilibrium dynamics,
provided $\mathbb{U}(t)$ in Eq.~\eqref{eq:UVdiag} depends on time and obeys
the equation $i\partial_t\mathbb{U}(t)=\mathbb{H}\, \mathbb{U}(t)$~\cite{mbeng2020quantum}.
The key point is that, in the ground-state and the dynamical cases, the state of the system keeps the Gaussian form
\begin{equation}
|\psi\rangle  
= {\calN} \; \exp{\bigg(\frac{1}{2} \sum_{j_1j_2=1}^L Z_{j_1j_2} \opcdag{j_1} \opcdag{j_2}\bigg) } \; |0\rangle ,
\label{eq:Gauss}
\end{equation}
where $Z=-(U^\dagger)^{-1}V^{\dagger}$ is a quadratic antisymmetric form [the $U$ and $V$ matrices being sub-blocks
  of $\mathbb{U}$, see Eq.~\eqref{eq:Umatr}],
$\calN$ a normalization factor, and $|0\rangle$ the vacuum of the $\hat c_j^{(\dagger)}$ fermions~\cite{mbeng2020quantum}.
This Gaussian form is preserved by the application of the exponential of any operator quadratic in $\hat c_j^{(\dagger)}$,
as for the Hamiltonian~\eqref{Eq:isingFermi}. In the next subsection we show that the two unravelings we are going
to consider amount precisely to the application to the state~\eqref{eq:Gauss} of exponentials of operators
quadratic in $\hat c_j^{(\dagger)}$. So the Gaussianity is preserved, together with the possibility
of a simple numerical treatment, whose complexity scales polynomially with $L$.
\subsection{The measurement process}\label{subsec:measurements}
The measurements of the environment give rise to a stochastic quantum dynamics. 
This means that the evolution of the state is provided by a trajectory $\ket{\psi_t}$ (also known as conditional state) that is the solution of a single realization of a stochastic process which models the quantum measurements. 
By ensemble averaging over the trajectories, we obtain the averaged density matrix 
\begin{equation}\label{mean:eqn}
  {\rho}(t) \equiv \overline{\ket{\psi_t}\bra{\psi_t}}\,,
\end{equation}
where $\overline{(\cdots)}$ marks the average over the stochastic ensemble of trajectories. Such density matrix follows a Lindblad-type evolution,~\footnote{Hereafter, unless specified, summations over the site indexes $j$ are
  intended to run over all the spins of the chain.}
\begin{equation}
  \frac{\ud}{\ud t} {\rho}(t) = - i \big[ \hat H, {\rho}(t) \, \big]
  - \frac{\gamma}{2} \sum_j \big[ \hat m_j \left[\hat m_j, {\rho}(t) \, \right] \big],
  \label{Eq:rho_medio}
\end{equation}
where $\hat m_j$ are the Hermitian measurement operators and $\gamma$ quantifies the strength of the coupling between
the system and the measurement apparatus. There are many stochastic processes giving the same Lindblad-type
evolution. 

We emphasize that the system we are considering is quantum, while the environment providing random measurements is a classical measurement apparatus. This situation differs from the quantum-open-system framework, where both the system and the environment are quantum~\cite{feynman,Cohen}.
  The classical measurement process is stochastic and each individual trajectory corresponds to a specific sequence of random measurement strokes~\cite{Daley2014,Plenio}.
The simple fact that the environment measurements have been performed ---even if we ignore the outcomes--- is enough to perturb the quantum evolution~\cite{Feynman_book}. The specific sequence of measurements and outcomes which has occurred provides the quantum trajectory corresponding to one realization of the stochastic process. If we were able to monitor when each click of the apparatus occurs, we would have access to the single trajectory. For this reason, the single trajectory actually has a physical meaning and can be experimentally observed in some context~\cite{siddiqi,PhysRevLett.57.1696,PhysRevLett.106.110502,devoret,Plenio}. Being the system observed by the environment, the trajectories do not interfere with each other and give rise at each time to a classical probability distribution of states~\cite{Feynman_book}. The average over this distribution provides the Lindblad-type evolution of Eq.~\eqref{Eq:rho_medio}, which captures the mean effect of this stochastic dynamics.

In what follows we consider two different
measurement processes that are the unraveling of the same Lindblad equation~\eqref{Eq:rho_medio}.
\subsubsection{Quantum-state-diffusion with continuous measurements}
We aim to measure the number of fermions on the $j$th site of the chain, hence we set $\hat m_j \equiv \hat n_j$,
where $\hat n_j = \hat c^\dagger_j \hat c_j$ denotes the number operator.
In this case, the Lindblad master equation is obtained by simply substituting $\hat m_j \mapsto \hat n_j$ in Eq.~\eqref{Eq:rho_medio}.

We start assuming the system to be {\em continuously} measured, i.e., the information
is continually extracted from it and the strength of the measurements is proportional
to a small time interval $\delta t$~\footnote{``Small'' means much smaller than the time scale of the resulting Lindblad dynamics, whose order of magnitude is $1/\gamma$ in Eq.~\eqref{Eq:rho_medio}~\cite{Plenio,Cohen}}.
This can be obtained, for instance, by integrating the Schr{\"o}dinger equation over $\delta t$;
Thanks to the Markov approximation, the resulting integrals of the couplings to the environment can be considered
as Gaussian random variables, uncorrelated at different times~\cite{Plenio}.
In this setting (namely, the quantum-state-diffusion model) we can write the measured dynamics
as a collection of Wiener processes resulting in a stochastic Schr\" odinger equation
[see Eq.~\eqref{Eq:QSD} in Appendix~\ref{App:CMD}]. 
Discretizing the time, the approximate evolution of the state over one step $\delta t$ of the dynamics is 
\begin{equation}
  \ket{\psi_{t+\delta t}} \simeq C \, e^{\sum_j [\delta W_t^j + (2\braket{\hat n_j}_t -1) \gamma \delta t] \hat n_j } e^{-i \hat H \delta t} \ket{\psi_t},
	\label{Eq:psi_cm}
\end{equation}
with $\delta W_t^j$ being normal distributed variables with zero mean and variance $\gamma \, \delta t$, and $\braket{\cdot}_t = \braket{\psi_t| \cdot |\psi_t}$. 
To a good approximation, for small enough $\delta t$ this Trotterized evolution faithfully describes the real dynamics~\cite{DeLuca2019} and preserves the Gaussianity of the state (for technical details, see Appendix~\ref{App:CMD}).
\subsubsection{Quantum jumps}
Another possibility we consider is an occasional, yet abrupt, measurement of the quantum state. This is what happens, for instance, to the electromagnetic field coupled to a photodetector.
Namely, at each time interval, there is a chance for the state to be measured (i.e., the detector clicks)
and projected, thus undergoing a so-called {\em quantum jump}~\cite{Daley2014}.
An interpretation of this process is given by rewriting Eq.~\eqref{Eq:rho_medio} as
\begin{equation}\label{lind:eqn}
  \frac{\ud}{\ud t}{{\rho}(t)} = -i\big( \hat{H}_\text{eff} \,{\rho(t)}
  - \rho(t) \, \hat{H}^\dagger_\text{eff} \big) + \gamma\sum_j \hat{m}_j\rho(t)\,\hat{m}_j ,
\end{equation}
with
\begin{equation}
  \hat{H}_\text{eff} = \hat{H} - i \frac{\gamma}{2} \sum_j \hat m_j^2\,.
  \label{Eq:ne}
\end{equation}
The dynamics can be thought of as a deterministic non-Hermitian evolution
driven by $\hat{H}_\text{eff}$ plus a stochastic part generated by the possibility of measuring $\hat m_j$.  
In order to preserve the Gaussian form of the state, we choose a slightly different form of the measurement operators
\begin{equation}
	\hat m_j \mapsto (\hat 1+\alpha \hat{n}_j),
	\label{Eq:l_gauss}
\end{equation}
with $\alpha>0$ real and $\hat 1$ being the identity operator. Substituting these operators in Eq.~\eqref{lind:eqn}, we easily see that the Lindblad master equation has the form of Eq.~\eqref{Eq:rho_medio} with $\hat m_j$ given by $\alpha \hat{n}_j$. So, for $\alpha=1$, the master equation is the same as in the quantum-state-diffusion case above, although the measurement operators are different.
Now, defining the quantity
\begin{equation}
  p_j  = \gamma \left[ 1 + \alpha(\alpha + 2)\braket{\hat{n}_j}_t \right]\,,
\end{equation}
and discretizing the time with intervals $\delta t$, the quantum-jump evolution $\ket{\psi_t} \to \ket{\psi_{t+\delta t}}$ can be obtained by applying at each step $\delta t$
\begin{enumerate}
\item with probability $\pi_j = p_j \, \delta t$, the jump operator
  \begin{equation}
    \ket{\psi_t}\to \frac{(\hat 1+\alpha\hat{n}_j)}{p_j} \ket{\psi_t} ;
    \label{eq:QJump}
  \end{equation}
\item with probability $p=1-\sum_j \pi_j$, the evolution operator associated to non-Hermitian Hamiltonian
  $\hat{H}_{\rm eff}=\hat{H} - \tfrac{ i\gamma\alpha(2+\alpha)}{2} \sum_j\hat{n}_j^2$:
  \begin{equation}
    \ket{\psi_t}\to\nep^{-i\hat{H}_{\rm eff}\delta t}\ket{\psi_{t}} .
  \end{equation}
\end{enumerate}
Obviously, operation $2.$ preserves the Gaussianity, since $\hat{H}_{\rm eff}$ is quadratic in the $\hat c_j^{(\dagger)}$ fermions of Sec.~\ref{subsec:Hamiltonian}. The same holds for operation $1.$, since the identity
\begin{equation}
  \hat 1+\alpha\hat{n}_j = e^{\log(1+\alpha)\hat{n}_j}
\end{equation}
guarantees that the operator applied to $\ket{\psi_t}$ in~\eqref{eq:QJump}
is the exponential of an operator which is quadratic in the $\hat c_j^{(\dagger)}$ fermions, thus preserving Gaussianity (see Appendix~\ref{App:qj}). 
%

\begin{figure}[!t]
  \includegraphics[width=0.42\textwidth]{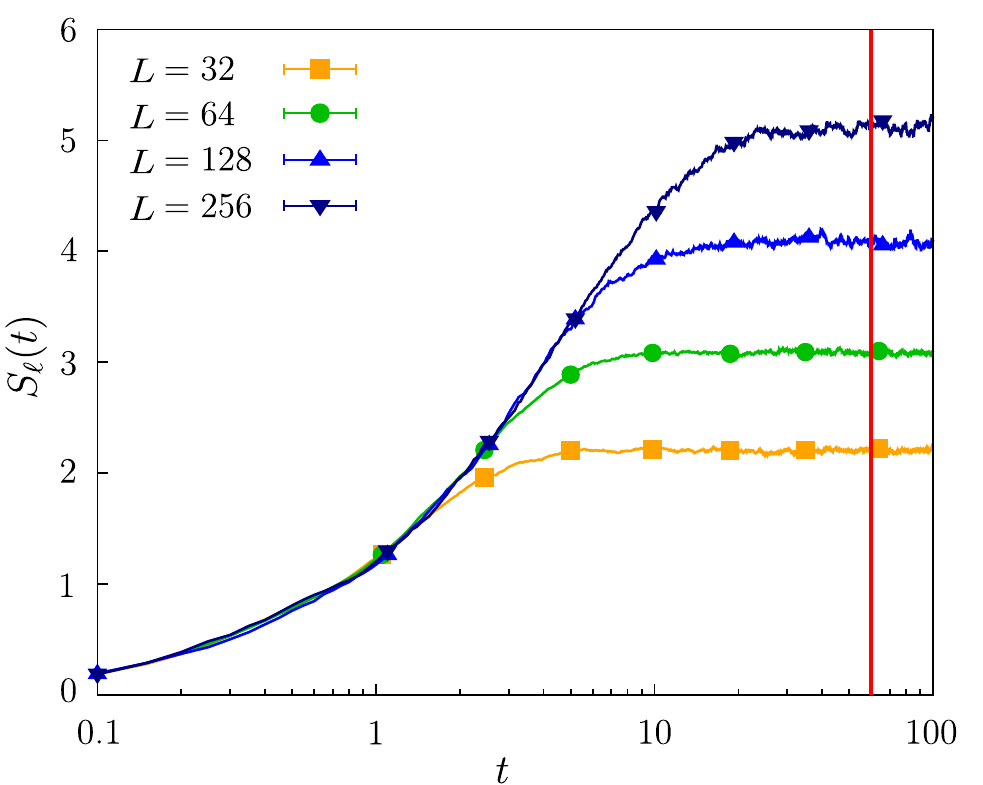}
  \caption{Ensemble-averaged entanglement entropy as a function of time during a continuous measurement process,
    for different system sizes $L$, with $\ell=L/4$, $h_f = 1.0$, and $\gamma = 1.0$.
    The red line at $t = t^\star = 60$ marks the time from which we evaluate the time-averaged entanglement entropy.
	Data are plotted in a semilog scale.}
   \label{Fig:ee_t}
\end{figure}

We conclude this section by remarking that the form of the quantum-jump operator in~\eqref{Eq:l_gauss} is quite general.
Indeed, due to the property $\hat{n}_j^2=\hat{n}_j$ holding for spinless fermions,
any operator of the form $\hat{O}_j=f(\hat{n}_j)$ can be written as $\hat{O}_j = [f(0)] \hat{1} + [f(1)-f(0)] \hat{n}_j$,
which is the same as~\eqref{Eq:l_gauss}, provided that $f(0)\neq 0$ and $f(0)\neq f(1)$.
\subsection{Entanglement entropy}\label{ent:sec}
We probe the entanglement properties of $\ket{\psi_t}$ in the different trajectories using the entanglement entropy, a widely used entanglement monotone~\cite{RevModPhys.81.865}. 
We divide the system in two partitions $A$ and $B$, so that the Hilbert space has a tensor-product structure $\mathcal{H}=\mathcal{H}_A\otimes\mathcal{H}_B$, and define the entanglement entropy of a state as the von Neumann entropy of the density matrix $\rho_A$ reduced to subsystem $A$:
\begin{equation} \label{entropy:eqn}
  S_{A}(\ket{\psi_t}\!\bra{\psi_t})\!=\!-\tr_{A}[{\rho}_{A}\ln{\rho}_{A}]\;\;{\rm with}\;\;{\rho}_A\!=\!\tr_B[\ket{\psi_t}\!\bra{\psi_t}]\,,
\end{equation}
where $\tr_B$ is the partial trace over $\mathcal{H}_B$.  
The entanglement entropy quantifies the mixedness of the reduced density matrix ${\rho}_A$, thus capturing correlations, since the reduced density matrix is mixed if the full system is correlated. When the full system is in a pure state, correlations are only quantum, therefore the entanglement entropy quantifies entanglement. In contrast, for global mixed states, correlations are both classical and quantum: In order to spotlight genuine quantum correlations one must resort to other quantities, like the two-qubit concurrence~\cite{wootters} or the quantum discord~\cite{rossini}, which are notoriously known to be hardly accessible for systems with more than two or three qubits.
 
In our case we have a random process giving rise to a statistical ensemble of quantum trajectories. This means that, at any time $t$, we have a distribution of pure states. This distribution is classical, there being no interference between the states in different trajectories~\cite{Feynman_book}. In order to evaluate entanglement at some time $t$, we first evaluate the entanglement entropy over the single pure-state realizations $\ket{\psi_t}$, and then average over the distribution. The two operations do not commute, due to the fact that the entanglement entropy is a nonlinear function of $\ket{\psi_t}\!\bra{\psi_t}$, and one has 
\begin{equation}
  \overline{S_A(\ket{\psi_t}\bra{\psi_t})}\leq S_A(\overline{\ket{\psi_t}\bra{\psi_t}}) \,.
\end{equation}
The second term of this inequality is larger because also classical correlations arising from the average over the classical distribution of states appear in the system. We are interested only in the quantum correlations, the ones leading to entanglement, so we consider $\overline{S_A(\ket{\psi_t}\bra{\psi_t})}$.

We emphasize that for quantities linear in $\ket{\psi_t}\!\bra{\psi_t}$, like the expectation values of observables $\hat{A}$, everything commutes and one has $\overline{\braket{\psi_t|\hat{A}|\psi_t}}=\tr[\hat{A}\,{\rho}(t)]$, where ${\rho}(t)=\overline{\ket{\psi_t}\bra{\psi_t}}$ is the averaged density matrix in Eq.~\eqref{mean:eqn}. So, linear quantities depend only on the averaged density matrix obeying the Lindblad equation and are not able to disclose differences between the various unravelings. In order to see such differences, one must resort to nonlinear quantities. In the main text we focus on the entanglement entropy and in Appendix~\ref{entone:sec} we show that other two of such quantities ---the second R\'enyi entropy and the square equal-time correlation function--- are able to detect the entanglement transition occurring in the quantum-state-diffusion.

For the evaluation of the entanglement entropy, we choose as subsystem $A$ a $\ell$-site long subchain, with $\ell=L/4$. For simplicity of notation we write $S_l(t)\equiv\overline{S_{A}(\ket{\psi_t}\!\bra{\psi_t})}$. Due to the Gaussian nature of the state, we can evaluate the entanglement entropy in a simple way, from the computational point of view (see Appendix~\ref{subsec:entanglement}).

\section{Quantum-state-diffusion: results}
\label{Sec:Rqsd}

We now present the results obtained under the assumption that the system is continuously monitored.
As detailed below, the entanglement entropy $S_\ell(L)$ of a subsystem of size $\ell$ undergoes a measurement-induced
transition from subextensive to area-law growth with the global system size $L$;
the crossover point strongly depends on the system parameters.

We prepare the system in the ground-state of the Ising Hamiltonian~\eqref{Eq:ising}
with a transverse field $h_i = +\infty$, and then we quench it to a value $h_f$ finite.
We monitor the dynamics of $S_\ell(t)$ for a subsystem of length $\ell = L/4$~\footnote{We have chosen $\ell=L/4$
  for computational convenience. However we carefully checked that, in the thermodynamic limit,
  our results qualitatively depend only on the system size and not on the subsystem one}.
$S_\ell(t)$ is a nonlinear function of the reduced density matrix, as discussed in Appendix~\ref{subsec:entanglement};
we evaluate it over $N_\text{rand}$ stochastic trajectories and then perform an ensemble average.
For the rest of this paper, we set $N_\text{rand} = 10^2$ (except when measuring the
correlation function in Fig.~\ref{Fig:corr}, where we set $N_\text{rand} = 3 \times 10^2$)
and, for quantum-state-diffusion, we take an integration step $\delta t = 0.05$. Such values of $N_\text{rand}$ and $\delta t$ have been selected after careful testing that
  the numerical errors introduced by these cutoffs are not affecting the results, on the scales
  of the figures presented in this work. We explicitly checked that, by increasing $N_\text{rand}$ over an order of magnitude, the results remain stable. Since we consider global quantities, where an average over sites is implicit, fluctuations average out and $N_\text{rand} = 10^2$ is enough to reach good convergence.

In Fig.~\ref{Fig:ee_t} we show some prototypical trajectories of the entanglement entropy in time.
The various colors refer to different system sizes, as indicated in the legend. The data have been taken
with $h_f = 1$ and $\gamma = 1$, but the qualitative behavior is not affected by this specific choice:
\begin{figure}[!t]
  \includegraphics[width=0.45\textwidth]{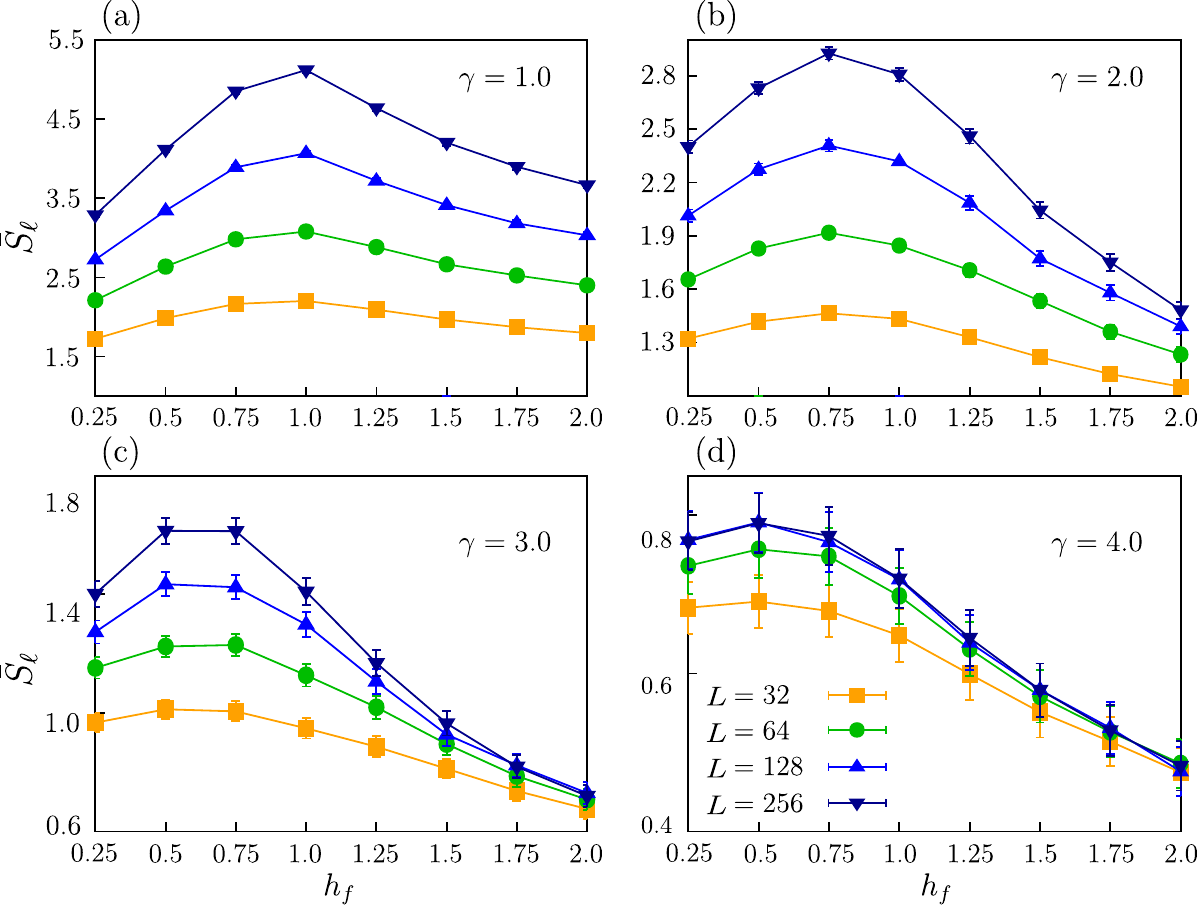}
  \caption{Asymptotic entanglement entropy $\bar{S}_\ell$ (obtained by averaging over times
    larger than $t^\star = 60$---see red line in Fig.~\ref{Fig:ee_t}) as a function of the post-quench
    transverse field $h_f$ for the measurement rates $\gamma= 1, 2, 3, 4$ [panels (a), (b), (c), (d), respectively].
    The various colors correspond to different system sizes.}
  \label{Fig:ee_tdf}
\end{figure}
After a transient time $t^\star(h_f, \gamma, L)$ that depends on the quench amplitude, the measurement rate,
and the system size, the entanglement entropy reaches an asymptotic value that may obey a subvolume or an area-law behavior.
We assume $t^\star = 60$ (red line in Fig.~\ref{Fig:ee_t}), after a careful a-posteriori check that, for $t > t^\star$,
all the trajectories have converged to the asymptotic value.

\begin{figure*}[!t]
  \includegraphics[width=0.98\textwidth]{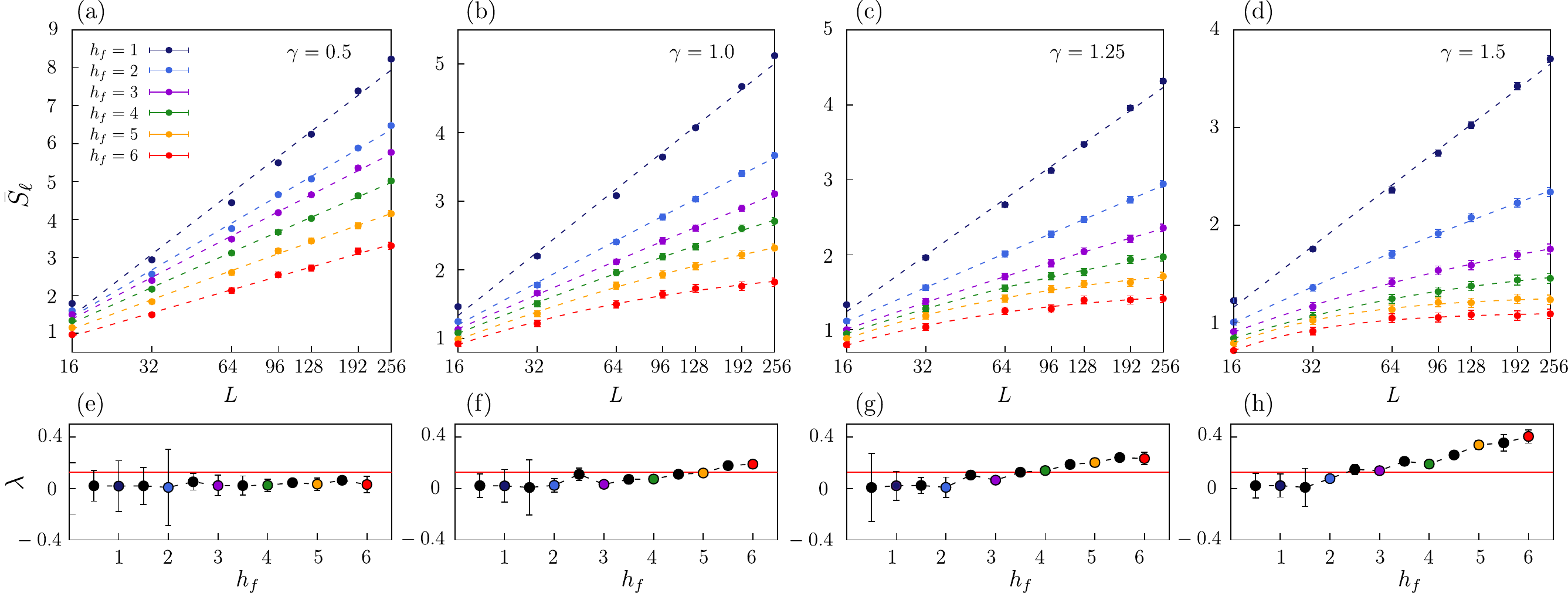}
  \caption{Top panels: Asymptotic entanglement entropy $\bar{S}_\ell$ as a function of the system size $L$
    (in semilog scale) for several values of $h_f$ (different colors), at fixed measurement
    rates $\gamma= 0.5, 1, 1.25, 1.5$ [panels (a), (b), (c), (d), respectively].
    We identify two behaviors: either $\bar{S}_\ell$ grows logarithmically with $L$,
    or it settles on a constant value.
    The crossover between these two regimes locates a measurement-induced phase transition.
    Dashed lines are obtained by fitting the data up to the largest available size, $L_\text{max} = 256$,
    with the function in Eq.~\eqref{Eq:Fit_Tanh}.
    Bottom panels: $\lambda$ as a function of $h_f$, for $\gamma= 0.5, 1.0, 1.25, 1.5$
    [panels (e), (f), (g), (h), respectively]. The color of the points refers to the corresponding entanglement
    trajectory in the top panels. The red line highlights the value
    $1/\log L_\text{max}$, corresponding to the validity bound
	of the fit: anytime $\lambda \log L_\text{max}$ is of order 1 or larger,
        we consider the fit reliable and the curve follows a
    hyperbolic tangent behavior (i.e., the entanglement shows an area-law behavior). In contrast, for $\lambda \log L_\text{max}\ll 1$ the behavior is indistinguishable from a straight line marking a subextensive scaling of the entanglement.}
  \label{Fig:tanh}
\end{figure*}

We define the asymptotic entanglement entropy
\begin{equation} \label{timav:eqn}
  \bar{S}_\ell =  \frac{1}{T-t^\star} \int_{t^\star}^T dt\ S_\ell(t)
\end{equation}
as the long-time-averaged entanglement entropy, where $t^\star$ is the transient time and $T$ the total simulation time, chosen long enough that convergence is reached.
The behavior of $\bar{S}_\ell$ is reported in Fig.~\ref{Fig:ee_tdf}, as a function of $h_f$ and for different
measurement rates $\gamma$. For small values of $\gamma$, the curves at different system sizes are well separated,
thus suggesting a size-dependence of the entanglement entropy. As we will show later, this dependence agrees with
the conformal scaling $\bar{S}_\ell(L) \sim \log(L)$. For larger $\gamma$ values, even though the subextensive behavior
survives when considering small system sizes and small transverse fields, a collapse of the curves emerges.
We emphasize that the peak at $h_f \approx 1$, reminiscent of the unitary dynamics 
discussed in Sec. \ref{subsec:Hamiltonian}, appears for small values of $\gamma$ in correspondence of the quantum critical point.
Because of the competition between the Hamiltonian evolution and the non-conserving one,
when increasing the measurement rate $\gamma$, this peak progressively shifts toward smaller transverse fields and eventually disappears. 
We remark that a similar behavior is observed in the ground-state entanglement of the quantum Ising chain at finite temperature~\cite{Amico2007}.   

Figure~\ref{Fig:tanh} displays the asymptotic entanglement entropy $\bar S_\ell$ as a function of the system size $L$
(notice the logarithmic scale on the $x$ axis), for different values of $\gamma$ (cf. different panels)
and $h_f$ (cf. different colors). We distinguish two behaviors: Some trajectories show a logarithmic growth $\bar S_\ell$ with $L$.
In this case, the data points follow a linear fit, whose slope determines the central charge of the associated
conformal description. When increasing $h_f$, the trajectories bend to eventually settle on a constant value,
i.e., an area-law behavior. The bending point, corresponding to the crossover point from the subextensive regime
to the area-law one, strongly depends on $h_f$, $\gamma$ and $L$.
We define $h_f^c$ as the critical transverse field where the crossover between area-law and logarithm-law takes place. 

The dashed lines in the top panels of Fig.~\ref{Fig:tanh} are obtained by fitting the data up to $L_\text{max} = 256$ with
\begin{equation}
	f(L) \sim\tanh \big[ \lambda(h_f, \gamma) \log(L) \big].
  \label{Eq:Fit_Tanh}
\end{equation}
In the bottom panels of Fig.~\ref{Fig:tanh} we show $\lambda(h_f, \gamma)$ as a function of $h_f$, for fixed $\gamma$.
The color of each point refers to that of the corresponding trajectory in the associated top panels.
Anytime the fit is reliable, the curve follows an hyperbolic tangent behavior, meaning that
the entanglement entropy will eventually attain an area-law regime. The fit is not reliable when $\lambda L_{\rm max}\ll 1$ and the curve is indistinguishable from a straight line, suggesting a logarithm-law behavior.
In practice, we consider the fit to be reliable whenever $\lambda(h_f,\gamma) \, \log L_\text{max}$ is of order 1
or larger. Therefore we take the relation
\begin{equation}
  \lambda(h_f^c,\gamma) \log L_\text{max} \sim 1
\end{equation}
as a qualitative estimate of the crossover transverse field $h_f^c$ for the given $\gamma$ [this condition is marked by the red horizontal lines in Fig.~\ref{Fig:tanh}(e-h)].
Hereafter, for simplicity of notation, we set $\lambda(h_f, \gamma) \equiv \lambda$. 
%

In Fig.~\ref{Fig:tdf} we sketch the phase diagram obtained by carrying out the above analysis for two values of
$L_\text{max} = 192$ (small empty circles) and $L_\text{max} = 256$ (big filled circles)~\footnote{Unfortunately,
  our computational resources did not allow us to go beyond $L_\text{max}=256$,
therefore we cannot establish whether and how $h_f^c$ would change for larger
values of $L_\text{max}$ and, ultimately, in the thermodynamic limit.}.
We identify three regions in the phase diagram corresponding to an extensive, a subextensive, and an area-law regime.
For $\gamma = 0$, the dynamics is unitary and the asymptotic entanglement entropy obeys a volume-law scaling.
For $\gamma \ge 4$, the entanglement entropy follows an area-law, as predicted in Ref.~\cite{Turkeshi2021}.
For $0 < \gamma < 4$, there exists a critical line $h_f^c(L_\text{max}, \gamma)$ dividing a region
where the entanglement grows subextensively from another region where it exhibits an area-law scaling.
We notice that the smaller the measurement rate $\gamma$, the higher the critical transverse field $h_f^c$.
Moreover, at fixed $\gamma$, the critical transverse field reduces while increasing the system size.
In principle, this leaves open the possibility that, in the thermodynamic limit, the subextensive region
might eventually fade away, despite a numerical proof of this conjecture appears out of reach.

\section{Quantum jumps: results}
\label{Sec:Rqj}

We now compare the results in the previous section with those obtained by using a quantum-jump protocol.
We put the emphasis on some discrepancies between the results obtained with the two unravelings (Sec.~\ref{Subsec:jumps}).
Finally, we compare the results of the stochastic evolutions with those coming from non-Hermitian Hamiltonian evolution, showing that this approximation does not capture many features
of the entanglement entropy (Sec.~\ref{Subsec:noclick}).

\subsection{Quantum-jumps evolution}
\label{Subsec:jumps}

Let us consider the quantum-jump dynamics described in Sec.~\ref{subsec:measurements}.
We discretize the time in steps separated by $\delta t$ and, at each time step, we extract a random number $r\in [0, 1]$.
If $r > \sum_j \pi_j $ we do not perform any measure; otherwise, if $\pi_j<r\leq \pi_{j+1}$ we measure $\hat m_j$
according to Eq.~\eqref{eq:QJump}. 
Attention must be paid in the choice of $\delta t$: In fact, if the time step is too large, the probabilities $\pi_j$
might exceed one. After a convergence check, we fixed the time step $\delta t^{-1} \propto 8 N \gamma\alpha$.

\begin{figure}[!t]
  \includegraphics[width=0.45\textwidth]{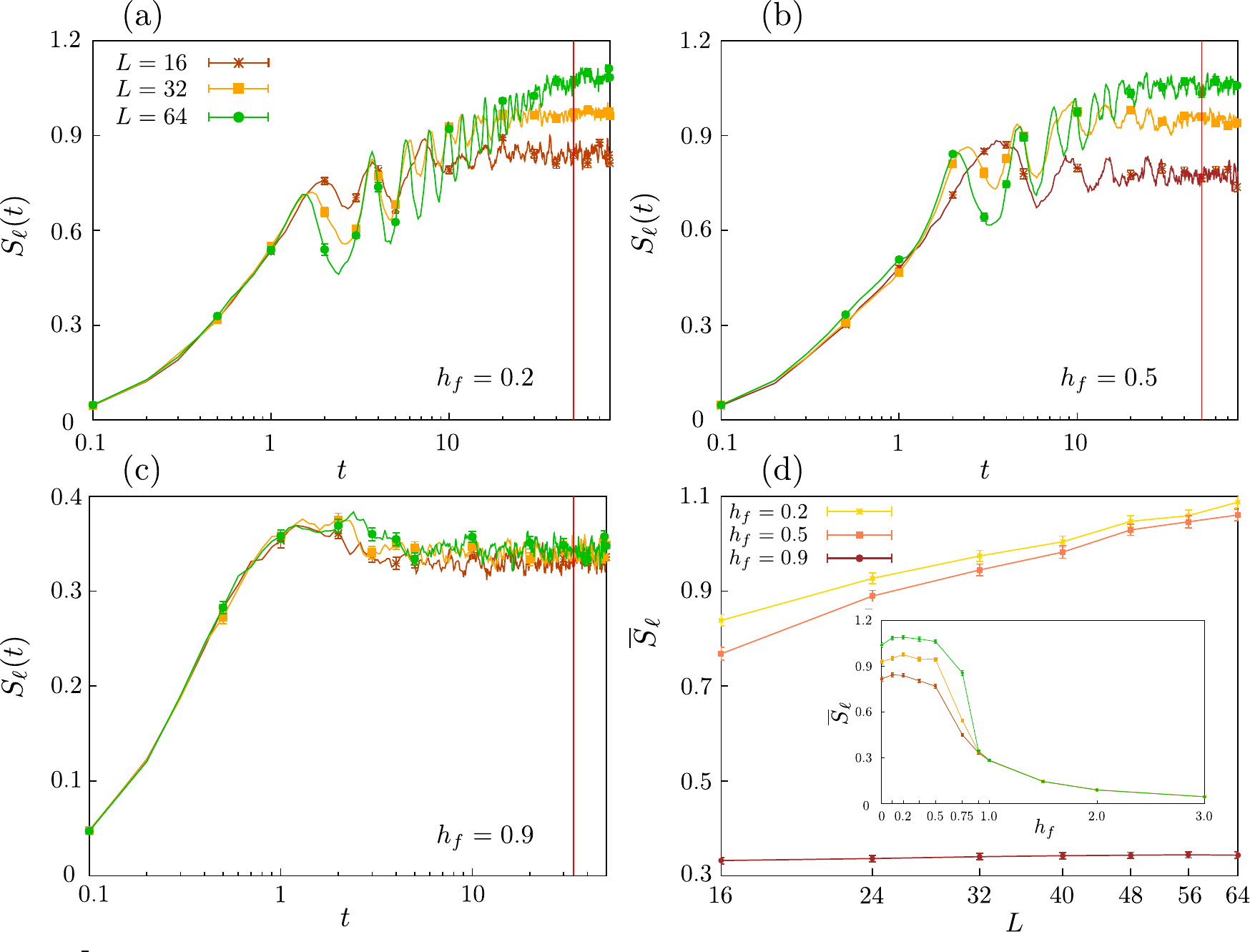}
	\caption{Panels (a), (b), (c): The quantum-jump-dynamics entanglement entropy $S_\ell(t)$ versus time, for $\alpha = 1$, for $h_f = 0.2, 0.5, 0.9$ $\gamma = 0.5$.
    The various colors correspond to different system sizes $L$. Panel (d): The asymptotic value $\bar{S}_\ell$,
    obtained by averaging on times larger than $t^\star$ [cf. red line in panels (a), (b), (c)],
	for the three considered values of $h_f$, versus $L$ (in semilog scale). The inset of panel (d) shows $\overline{S}_\ell$ versus $h_f$ evaluated at $L = 16, 32, 64$.}
  \label{Fig:ee_qj}
\end{figure}

Although the dynamics is different from the quantum-state-diffusion one,
we fix $\alpha = 1$ in order to recover the same master equation for the averaged density matrix.
In Fig.~\ref{Fig:ee_qj} we show the entanglement entropy trajectories for $\gamma = 0.5$, $h_f = 0.5, 1.0, 2.0$,
and various system sizes [panels (a), (b), (c), respectively].
Despite the differences in the asymptotic value, the entanglement entropy in time follows
similar trajectories as those in Fig.~\ref{Fig:ee_t}.
In Fig.~\ref{Fig:ee_qj}(d) we plot the value of the asymptotic entanglement entropy $\bar S_\ell$ versus the system size $L$, showing that $\bar S_\ell$ is experiencing a transition from a logarithmic-scaling phase towards an area-law one.
The inset of Fig.~\ref{Fig:ee_qj}(d) shows the asymptotic entanglement entropy versus the final transverse field for different system sizes. We observe that, at fixed $L$ and for $h_f$ in the ferromagnetic phase, $\bar S_\ell$ increases with the system size. In contrast, for $h_f$ above a threshold close to $h_c$, $\bar S_\ell$ is constant with the system size. Therefore, we cannot exclude that this effect might be related to the ground-state critical point $h_c$.

\begin{figure}[!t]
  \includegraphics[width=0.42\textwidth]{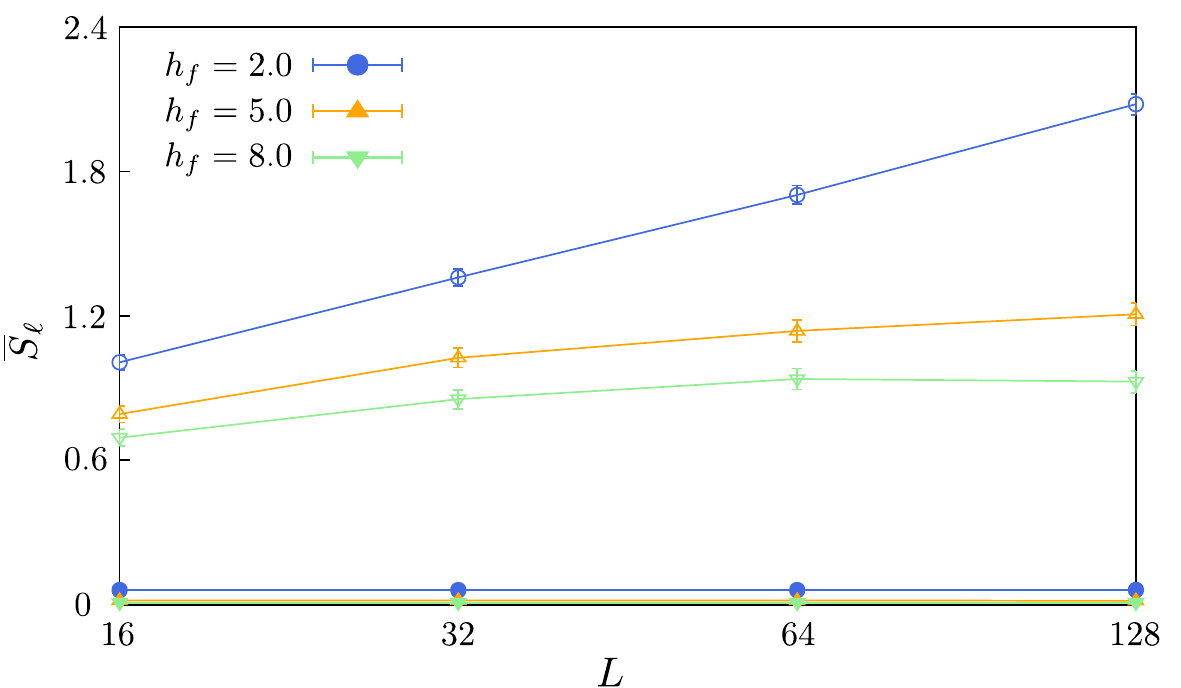}
	\caption{Filled dots: Asymptotic entanglement entropy $\bar{S}_\ell$ versus the system size in the quantum-jump dynamics,
    for $h_f = 2, 5, 8$, and $\gamma = 1.5$ (blue, orange, and light-green circles in Fig.~\ref{Fig:tdf}, respectively).
	Empty dots: the corresponding results obtained with the quantum-state-diffusion scheme.
	Note the stable area-law behavior of the quantum-jumps dynamics for $h_f=2.0$, where, in contrast, quantum-state-diffusion shows a logarithmic-law (blue curves).
	Errors are comparable with the symbols size size (not shown).
      Data are plotted in semilog scale.}
    \label{Fig:ee_g4}
\end{figure}

Despite giving rise to the same Lindbladian, the quantum-jump dynamics does not provide the same phase diagram
as the quantum-state-diffusion protocol (Fig.~\ref{Fig:tdf}). 
The former, in fact, gives rise to a region of logarithmic scaling of the entanglement enrtopy smaller than that of the quantum-state-diffusion dynamics.
To better clarify this point, we fix $\gamma = 1.5$ and pick up three representative points of the quantum-state-diffusion phase diagram,
in the subextensive region ($h_f = 2.0$ -- blue circle in Fig.~\ref{Fig:tdf}), 
on the crossover line ($h_f = 5.0$ -- orange circle in Fig.~\ref{Fig:tdf}),
and in the area-law phase ($h_f = 8.0$ -- light-green circle), and we plot the asymptotic entropy versus $L$ in~\ref{Fig:ee_g4}.
All the curves obey an area-law scaling. 
As a comparison, we also plot the same curves obtained with the quantum-state-diffusion approximations (empty points and dashed lines). We see that the curve for $h_f=2.0$ shows a logarithmic scaling, in contrast with the corresponding area-law behavior of the quantum-jump protocol. For both protocols the curves for $h_f=5.0, 8.0$ display an area-law behavior. In all the considered cases, the entanglement entropy for the quantum-state-diffusion is quite larger than the quantum-jump protocol. 

\subsection{Non-Hermitian Hamiltonian approximation}
\label{Subsec:noclick}

Finally, we consider the entanglement dynamics in the so-called {\em no-click limit}, meaning that the entanglement entropy
is evaluated over post-selected trajectories that did not jump during the evolution.
This class of trajectories follows a deterministic dynamics driven by the non-Hermitian Hamiltonian in Eq.~\eqref{Eq:ne}. 
There are some results in the literature showing that sometimes it is possible to find a correspondence
between the measurement-induced entanglement transition and the purely non-Hermitian
dynamics~\cite{Turkeshi2021,biella2021,sarang2021,PhysRevE.104.034107}.

As an example, the authors of Ref.~\cite{Turkeshi2021} show that the entanglement transition of the Ising chain
in the absence of transverse field ($h = 0$) can be quantitatively located also by looking at the non-Hermitian dynamics
of the entanglement entropy. This result is corroborated by some observations on the spectrum of Eq.~\eqref{Eq:ne},
\begin{equation}\label{lacca:eqn}
	\omega_k^\text{n-H}\Big(h = \frac{i\gamma}{4}\Big) = 2J\sqrt{1 - \frac{\gamma^2}{16 J^2} - i \frac{\gamma}{2J}\cos k}
\end{equation}
obtained by substituting $h = i \gamma/4$ in Eq.~\eqref{Eq:eigv}.
In fact, the critical rate $\gamma_c = 4$ is the one at which the real part of the spectrum vanishes
and the imaginary part becomes gapped. 
Despite this, the non-Hermitian evolution fails to predict the entanglement entropy dynamics, as soon as
a finite transverse field is considered.
To prove this, we simulate the non-Hermitian dynamics for quenches towards $h_f > 0$.
In order to have the same non-Hermitian Hamiltonian of Ref.~\cite{Turkeshi2021} and to simplify the comparison,
for these simulations we set $\alpha(2+\alpha)=1$, i.e., $\alpha =\sqrt{2}-1 \approx 0.41$.
We point out that, even though the Lindblad equation for the averaged density matrix with $\alpha \sim 0.41$ is different from that considered in Subsec.~\ref{subsec:measurements}, in the no-click limit the evolution of the state is fully determined by the non-Hermitian dynamics. We checked the non-Hermitian evolution with $\alpha = 1.0$ without finding qualitative differences with the other case.

\begin{figure}[!t]
  \includegraphics[width=0.45\textwidth]{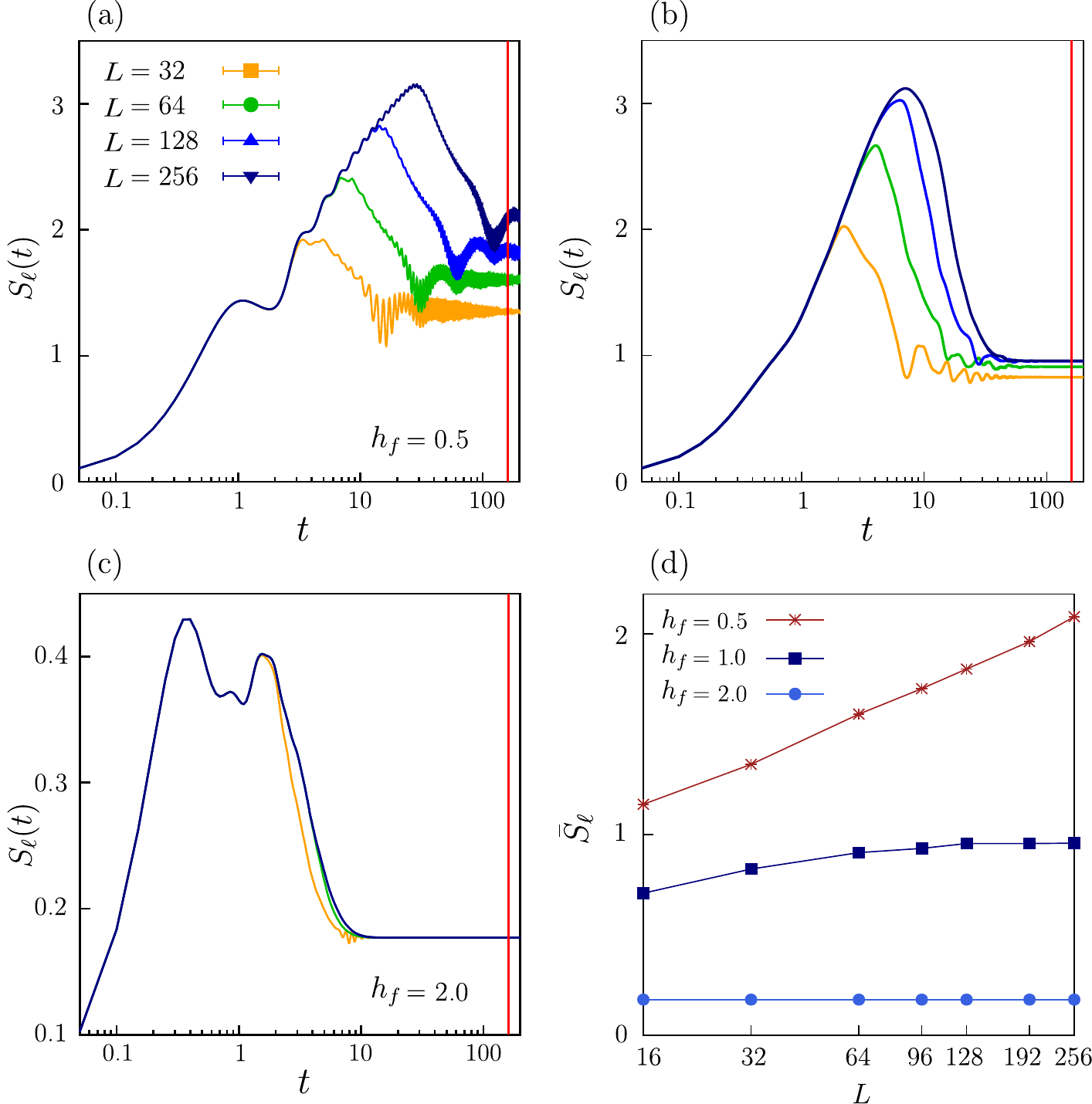}
  \caption{Panels (a), (b), (c): Non-Hermitian evolution of the entanglement entropy $S_\ell(t)$ as a function of time,
    for $h_f = 0.5, 1.0, 2.0$ [panels (a), (b), (c), respectively], and $\gamma = 0.5$.
    The various colors correspond to different system sizes. Panel (d) shows the asymptotic value $\bar{S}_\ell$
    [obtained by averaging on times larger than $t^\star = 160$ ---the red line in panels (a), (b), (c)]
    for the three $h_f$ as a function of $L$.
    The curves in panel (a) show a larger transient than the others, therefore, before evaluating the $\bar{S}_\ell$,
    we carefully checked for their behavior at later times.}
  \label{Fig:ee_nh}
\end{figure}

In Figs.~\ref{Fig:ee_nh}(a), ~\ref{Fig:ee_nh}(b), and ~\ref{Fig:ee_nh}(c) we show non-Hermitian evolution trajectories with the same parameters of Figs.~\ref{Fig:ee_qj}(a), \ref{Fig:ee_qj}(b) and~\ref{Fig:ee_qj}(c), respectively.
We notice that not only the entanglement follows qualitatively different 
time traces, but also the convergence is much slower than in the quantum-jump case.
In Fig.~\ref{Fig:ee_nh}(d) we show the asymptotic entanglement entropy as a function of the system size for the three values of $h_f$ considered so far. What emerges is that this approximate dynamics predicts a different entanglement scaling than the quantum-state-diffusion and the quantum jumps. For instance, while the curve at $h_f = 2.0$ in Fig.~\ref{Fig:ee_nh}(d) is constant in $L$, the same curve obtained with the quantum-state-diffusion dynamics at the same $\gamma$ grows logarithmically with $L$ [Fig.~\ref{Fig:tanh}(a)].
Moreover, the asymptotic entanglement entropy obtained from the non-Hermitian evolution is monotonous decreasing with $h_f$. As a consequence, we do not find any evidence of the critical point for small measurement rates, in contrast with the quantum-state-diffusion model [see Fig.~\ref{Fig:ee_tdf}].
Interestingly, the considerations done on the spectrum of the non-Hermitian Hamiltonian at $h_f = 0$~\cite{Turkeshi2021}
cannot be extended to the case of a complex transverse field, namely $h_f > 0$. In fact, as it emerges by substituting
$h = h_f + i\gamma/4$ in Eq.~\eqref{Eq:eigv}, it is not possible to find a $k$-independent $\gamma$ that makes
the real part of the spectrum vanishing, while keeping the imaginary part gapless.
    
\section{Discussion and conclusion}
\label{sec:conclusions}
We focused on measurement-induced entanglement phase transitions in the quantum Ising chain, subject to different
measurement processes resulting in the same Lindblad master equation for the averaged density matrix. 
We chose two different unraveling modeling measurements: (i) the quantum-state-diffusion model, occurring weakly
but continuously in time, and (ii) quantum-jump description, occurring abruptly but randomly in time.
In doing this, to allow simulations for larger systems, we payed attention in choosing measurement operators
which preserve the Gaussianity of the evolving quantum state.   

From one side we found is that the quantum-state-diffusion dynamics predicts a crossover form a subextensive (logarithm-law) phase
of the entanglement entropy to an area-law one. From the other side, the location of this crossover moves to different parameters when considering a quantum-jump dynamics, suggesting a smaller subextensive region, although the Lindblad equation is the same.
As expected, we never recover the volume-law growth typical of the unitary evolution.

Our results emphasize the fact that the entanglement entropy is determined by the quantum correlations
contained in the single quantum trajectory, that strongly depend on the choice of the unraveling. During the averaging process leading to the Lindblad equation, 
classical correlations appear as well. The resulting correlations in the averaged density matrix are of both types and cannot be disentangled from each other. Different unravelings provide different amounts of quantum correlations in the single quantum trajectories ---and then different behaviors of the entropy--- although the average density matrix is always the same.

In this framework, different behaviors of entanglement come from the competition between unitary dynamics and measurement process. According to the environment measurement process over which the quantum-trajectory averaging is performed,
the same Lindblad equation ---and the same average density matrix--- can come from a ``more quantum'' evolution,
where the constructive effect of the unitary dynamics leads to large entanglement, or from a ``more classical'' one,
where the destructive effect of the measurements prevails. The same competition gives rise to the transitions between different entanglement dynamical phases, when the unraveling is fixed and the coupling with the environment is varied (as occurs in the quantum-state-diffusion case).

Even though, to the best of our knowledge, for integrable fermionic models there are no examples
of measurements preserving the volume-law scaling, some results in literature propose more complicated models displaying it,
as for the case of measurements performed at discrete periodic times in ergodic phase~\cite{Skinner2019}
or in integrable MBL systems, provided appropriate measurement operators are chosen~\cite{Lunt2021}. 
It would be tempting to check whether it is possible to find a set of continuously measured operators
which can provide, in our framework, a volume-law asymptotic entanglement entropy.
In this respect, an interesting possibility would be to exploit, for instance, long-range interactions
in the definition of the measurement operators~\cite{sierant2021dissipative}. 

Finally, we comment on the experimental relevance of the results presented above. In fact, even though we are not aware of a genuine detection of the entanglement entropy, there are some experimental works which discuss the possibility of measuring the R\'enyi entropy~\cite{islam2015, Brydges2019}, another entanglement monotone that is expected to capture the entanglement transitions (for a more quantitative discussion, see Appendix~\ref{a:reny}). Interestingly, some recent experimental proposals for measuring the entanglement in measurement-driven dynamics similar to ours have been put forward~\cite{czischek2021} and realized~\cite{noel2021}. Moreover, as discussed in Appendix~\ref{a:corr}, information on the entanglement can be inferred also by looking at the asymptotic correlations, suggesting a different way to detect such transitions.
Further experimental advances in this direction could help to open new directions in this field,
shading some light on the problem of the unraveling dependence of entanglement transitions. 

{\em Note added:} During the completion of this work, we became aware of a related manuscript~\cite{Turkeshi2021b}
discussing the emergence of entanglement transitions in free-fermion models evolving under quantum jumps.

\acknowledgments

We thank A.~Biella, S.~Diehl, R.~Fazio, R.~Khasseh, P.~Lucignano, G.~Passarelli, and V.~Russomanno for useful discussions and comments.
We acknowledge support from the Italian MIUR through PRIN Project No. 2017E44HRF. We thank M. Schirò adn X. Turkeshi for pointing out a possible mistake in the first version of our paper, thus prompting us to double check our previous results.
The simulations have been carried using Armadillo c++ library~\cite{armadillo1, armadillo2}.

\appendix
\section{Entanglement related quantities} \label{entone:sec}
In this appendix we provide some examples of other entanglement related quantities to explore. First we discuss the possibility of detecting entanglement transitions by looking at the second-order R\'enyi entropy and then we consider the equal-time square correlations.

\subsection{Second-order R\'enyi entropy}\label{a:reny}
\begin{figure}
  \centering
  \includegraphics[width=0.37\textwidth]{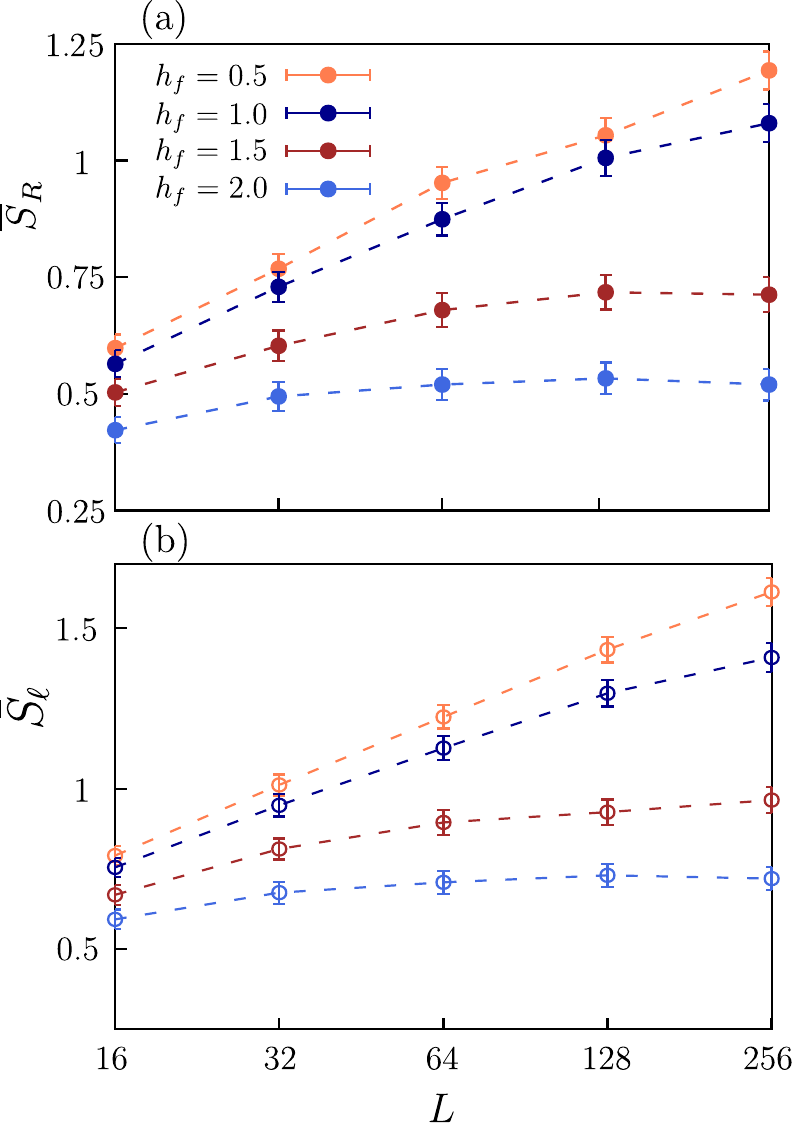}
  \caption{Asymptotic second-order R\'enyi entropy [Panel (a)] and entanglement entropy [Panel (b)]
      versus the system size for $\gamma = 3.0$ and $h = 0.5, \, 1.0, \, 1.5, \, 2.0$. 
      The two entropies are in good agreement. Data are plotted in semilog scale.}
  \label{Fig:reny}	
\end{figure}
The R\'enyi entropies are a class of entanglement monotones widely diffused in the quantum-information field~\cite{RevModPhys.81.865}. Given a pure state $\ket{\psi_t}$ and a partition of the system in two subsystems $A$ and $B$, the R\'enyi entropy of order $\beta$ is defined as 
\begin{equation} \label{renna:eqn}
  H_\beta(\ket{\psi_t}\bra{\psi_t}) = \frac{1}{1-\beta} \ \ln \tr_A \big[ \big({\rho}_A\big)^\beta \big] \,,
\end{equation}
where ${\rho}_A=\tr_B[\ket{\psi_t}\bra{\psi_t}]$ is the reduced density matrix of subsystem $A$ [see Eq.~\eqref{entropy:eqn}].
The R\'enyi entropy can be numerically evaluated in an efficient way for fermionic Gaussian states (see Appendix~\ref{subsec:entanglement}) and the nontrivial limit for $\beta \to 1$ corresponds to the von Neumann entropy discussed in the main text.
Of particular interest is the second-order ($\beta=2$) R\'enyi entropy 
\begin{equation}
  H_2(\ket{\psi_t}\bra{\psi_t}) = - \ln \tr \big[ \big({\rho}_A\big)^2 \big] \,.
\end{equation}
This quantity is experimentally achievable~\cite{islam2015, Brydges2019} and is related to the purity of the state, which has an important role in the study of decoherence~\cite{rossini}. As in the case of the entanglement entropy, we consider as the subsystem $A$ a $\ell$-site long subchain with $\ell=L/4$ and define $S_R(t)\equiv\overline{H_2(\ket{\psi_t}\bra{\psi_t})}$. We consider the time-average in the asymptotic state using a formula similar to Eq.~\eqref{timav:eqn}, $\bar{S}_R =  \frac{1}{T-t^\star} \int_{t^\star}^T dt \, S_R(t)$, with the total simulation time $T$ chosen long enough that convergence is reached. 

Since the R\'enyi entropy is an entanglement quantifier, we expect it to behave in a way similar to the entanglement entropy. Here we do not carry out an analysis as detailed as for the entanglement entropy, but we are only interested in checking that the behavior of the two quantities is similar. In particular, considering the quantum-state-diffusion unraveling (Sec.~\ref{Sec:Rqsd}), we expect to recover the same entanglement transition in its behavior. 
As an example, in Fig.~\ref{Fig:reny} we show the asymptotic time-averaged entanglement entropy $S_\ell$ and R\'enyi entropy $S_R$ versus the system size $L$, for different parameters. In fact the two quantities behave in a similar way, and when one displays an area-law scaling, the other follows the same scaling. 
This result suggests that the R\'enyi entropy reasonably behaves area-law in the same region where the von Neumann entropy does.

\subsection{Correlations}\label{a:corr}
Still focusing on the quantum-state-diffusion case, we can observe signatures of the entanglement transition also by looking at the square equal-time correlation. This quantity has been introduced in~\cite{Alberton2021} and is defined as
\begin{equation}
	C_j(t, r) = \overline{|\braket{\psi_t|\opcdag{j}\opc{j+r}|\psi_t}|^2}
\end{equation}
(note that in the limit $N_{\text{rand}}\to\infty$ there is no dependence on $j$, due to translation invariance). Intuitively, we may expect that a more entangled system is more correlated than a less entangled one. In particular, $C(t,r)$ should decay algebraically with $r$ in a state with subextensive entanglement, while it should decay exponentially when the entanglement entropy obeys an area-law scaling~\cite{Alberton2021}.
\begin{figure}[!t]
  \centering
  \includegraphics[width=0.45\textwidth]{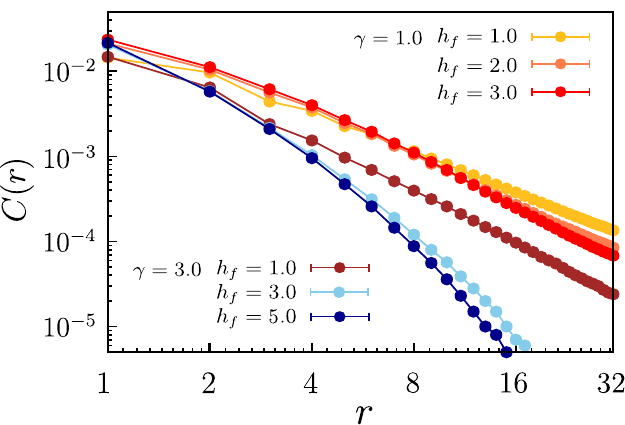}
  \caption{Asymptotic square equal-time correlation function $C(r)$ versus the site distance $r$. When the entanglement scales subextensively (curves with warm tones), the correlations display an algebraic decaying. When the entanglement scales as an area-law, the correlations seem to undergo a super-algebraic decay (curves with cool tones). Here we set $L = 128$ and averaged over $N_\text{rand}=300$ trajectories. Data are plotted in log-log scale, to better highlight power-law decays.}
  \label{Fig:corr}	
\end{figure}

The data reported in Fig.~\ref{Fig:corr} hint at this kind of behavior, where we show the asymptotic correlation
  \begin{equation}
    C(r) \equiv \frac{1}{L(T-t^\star)}\int_{t^\star}^T dt\,\sum_{j=1}^L C_j(t, r)
  \end{equation} 
versus the site distance $r$. The average over time in this formula is as in Eq.~\eqref{timav:eqn}, and we average also over sites in order to reduce the noise, and allow convergence of the average also for $N_{\text{rand}}$ not larger than $3\times 10^2$.
Comparing these results with those plotted in Fig.~\ref{Fig:ee_tdf}(a) and~\ref{Fig:ee_tdf}(c), we can make the following observations.
For parameters in the regime of subextensive entanglement scaling (i.e., curves with warm tones) we observe a slow power-law decay. In contrast, in the area-law regime for the entanglement (i.e., curves with cool tones) we notice a much faster drop which seems to considerably deviate from power-law.

\section{Diagonalization of the Ising model}\label{App:ising}
In this appendix we recall how to diagonalize the quantum Ising chain, described by the Hamiltonian in Eq.~\eqref{Eq:ising}.
First we introduce the Jordan-Wigner transformation,
\begin{equation}
  \hat \sigma^+_j = \big( e^{i \pi \sum_{\ell = 1}^{j-1} \hat n_{\ell}} \big) \, \hat c_j^\dagger ,
  \label{eq:JWT}
\end{equation}
where $\hat \sigma^\pm_j = \tfrac12 \big( \hat \sigma^x_j \pm i \hat \sigma^y_j \big)$
are the raising and lowering operators of the $j$th spin, 
that maps Eq.~\eqref{Eq:ising} into the spinless-fermion Hamiltonian
\begin{eqnarray}
  \hat H & = &- J \sum_{j = 1}^{L-1} (\hat c^\dagger_j \hat c_{j+1} + \hat c^\dagger_j \hat c^\dagger_{j+1} + {\rm h.c.})
  + h \sum_{j = 1}^L (2 \hat n_j - 1)\nonumber \\
  && + (-1)^N J (\hat c^\dagger_L \hat c_1 + \hat c^\dagger_L \hat c^\dagger_1 + {\rm h.c.}).
  \label{Eq:Ham_fermions}
\end{eqnarray}
In this expression, $\hat c_j^{(\dagger)}$ denote anticommuting fermionic annihilation (creation) operators,
$\hat n_j = \hat c^\dagger_j \hat c_j$ is the corresponding local number operator,
and $N = \sum_j \langle \hat c^\dagger_j \hat c_j \rangle$ is the total number of fermions.
The last term in Eq.~\eqref{Eq:Ham_fermions} accounts for the periodic boundary conditions and,
because of the highly non-local character of the transformation~\eqref{eq:JWT}, is strongly affected
by the parity sector in which one is working.
In our case we fix $N$ even, hence we assume antiperiodic boundary conditions in Eq.~\ref{Eq:Ham_fermions}.
The Hamiltonian~\eqref{Eq:Ham_fermions} can be written in the compact form
\begin{equation}
  \hat H = \tfrac{1}{2} \hat \Psi^\dagger \mathbb{H} \hat \Psi + \text{const.},
  \end{equation}
where
$\hat \Psi = (\hat c_1, \cdots, \hat c_L, \hat c_1^\dagger, \cdots, \hat c_L^\dagger)^T$
is the Nambu spinor introduced in the main text,
while
\begin{equation}
	\mathbb{H} = \begin{pmatrix} \ \ A & \ \  B \\ -B & - A \end{pmatrix}
		\label{Eq:bdg}
\end{equation}
is the so-called Bogoliubov-De Gennes Hamiltonian matrix,
with entries
\begin{equation}
  \begin{cases}
    A_{j,j} = h, \qquad A_{j, j+1} = \ \ A_{j+1, j} = -J/2, \\
    B_{j,j} = 0, \qquad B_{j, j+1} = -B_{j+1, j} = -J/2, \\
	A_{L,1} = A_{1,L} =	B_{L,1} = -B_{1,L} = (-1)^{N+1}J/2. \\ 
  \end{cases}
\end{equation}

We now define a set of new fermions $\hat \gamma_k$ ($k=1,\ldots,L$), obeying canonical anticommutation rules, as follows:
\begin{equation}
    \displaystyle \hat \gamma_k = \sum_{j=1}^L \big( U_{jk}^* \hat c_j + V_{jk}^* \hat c_j^\dagger \big). 
\end{equation}
The associated Nambu spinor is given by
$\hat \Phi =( \hat \gamma_1, \cdots, \hat \gamma_L, \hat \gamma_1^\dagger, \cdots, \hat \gamma_L^\dagger)^T = \mathbb{U}^\dagger\hat\Psi$,
with
\begin{equation}
  \mathbb{U} = \begin{pmatrix} U & V^* \\ V & U^* \end{pmatrix} .
  \label{eq:Umatr}
\end{equation}
The $\mathbb{U}$ matrix implements the so-called Bogoliubov transformation, which makes the Hamiltonian diagonal
in the $\gamma_k$ fermions [cf. Eq.~\eqref{eq:UVdiag}], with a dispersion relation given by Eq.~\eqref{Eq:eigv}.

\section{Continuous measurement dynamics}\label{App:CMD}
In this appendix we discuss more technical details of the dynamics in presence of continuous measurements. 

As stated in the main text, the dynamics of these kind of measurements is captured by a collection of Wiener processes.
A Wiener process is an ideal quantum walk with arbitrary small, independent, steps taken arbitrarily often,
that is normally distributed with zero mean and variance growing linearly in time.
The resulting stochastic Schr\" odinger equation reads 
\begin{eqnarray}
  d \ket{\psi_t} & = & - i \hat H dt \ket{\psi_t} + \bigg\{ \sum_j \sqrt{\gamma} (\hat n_j - \braket{\hat n_j} ) dW_t^j \bigg\} \ket{\psi_t} \nonumber \\
  && - \frac{1}{2} \bigg\{ \sum_j \gamma( \hat n_j - \braket{\hat n_j})^2 dt \bigg\} \ket{\psi_t},
  \label{Eq:QSD}
\end{eqnarray}
with $W_t^j$ independent Wiener processes. 
By Trotterizing Eq.~\eqref{Eq:QSD}, we obtain the approximate evolution in Eq.~\eqref{Eq:psi_cm}.

For a Gaussian state, this evolution reduces to that of the correlation matrices $U, V$ defined in Eq.~\eqref{eq:Umatr}.
In particular, it can be written as a two-step evolution driven first by the Hamiltonian part,
which is given by the unitary transformation
\begin{equation}
	\begin{bmatrix} U'(t + \delta t) \\ V'(t + \delta t) \end{bmatrix}
		= e^{-2i \, \mathbb{H} \, \delta t} \begin{bmatrix} U(t) \\ V(t) \end{bmatrix},
  \label{Eq:Ham_U}
\end{equation}
and then by the dissipative part
\begin{equation}
	\mathbb{W} = \exp \begin{bmatrix} T & 0 \\0 & -T\end{bmatrix} \begin{bmatrix} U'(t + \delta t) & (V'(t + \delta t))^* \\ V'(t + \delta t) & (U'(t + \delta t))^* \end{bmatrix},
\end{equation}
where $T$ is a $L \times L$ diagonal matrix defined as
\begin{equation}
  T_{jj} = \delta W_t^j + (2 \braket{\hat n_j}_t -1) \gamma \, \delta t.
\end{equation}
Since the dissipative part does not conserve the norm of the state, to keep it normalized we have to perform
a QR decomposition $\mathbb{W} = \mathbb{Q}\cdot \mathbb{R}$, with $\mathbb{Q}$ an orthogonal matrix
and $\mathbb{R}$ an upper triangular one. Thus the time-evolved state is simply $\mathbb{U}(t + \delta t) \equiv \mathbb{Q}$.~\cite{Turkeshi2021,DeLuca2019}

\section{Quantum jumps dynamics}\label{App:qj}
In this appendix we give some technical details on the quantum jump evolution described in the main text. 
This protocol is quite different from the continuous measurement one, since we choose $\hat m_j = \sqrt{\gamma}(\hat 1 + \alpha\hat{n}_j)$ as measurement operators.
By exploiting the operator identity
\begin{equation}
  \hat{n}_j=\frac{\nep^{x\hat{n}_j}-\hat 1}{\nep^x-1},
	\label{Eq:identity}
\end{equation}
it is easy to be convinced that these operators preserve the Gaussianity of the state. 
In fact, once fixed $x = \log(1 + \alpha)$, we have
\begin{equation}
  \hat{m}_j=\sqrt{\gamma}\left(\hat 1+\alpha\frac{\nep^{x\hat{n}_j}-\hat 1}{\nep^x-1}\right)=\sqrt{\gamma}\nep^{x\hat{n}_j}\,.
\end{equation}
This quantum-jump dynamics is well described by the stochastic equation
\begin{eqnarray}
	d \ket{\psi_t} & = & - i \hat H dt \ket{\psi_t} -\frac{\gamma}{2} \bigg\{ \sum_j (\hat m_j - \braket{\hat m_j} ) dt \bigg\} \ket{\psi_t} \nonumber \\
	&& + \bigg\{ \sum_j \left( \frac{\hat m_j}{\sqrt{\braket{\hat m_j}_t}} - 1\right)  dN_t^j \bigg\} \ket{\psi_t},
  \label{Eq:qj}
\end{eqnarray}
where $N_t$ are Poisson processes with $dN_t^j = 0,1$, $(dN_t^j)^2= dN_t^j$ and $\overline{dN_t^j} = \gamma dt \braket{n_j}_t$.
The evolution driven by the non-Hermitian Hamiltonian in Eq.~\eqref{Eq:ne} is obtained by solving the equation  
\begin{equation}
  i \frac{d}{dt} \begin{bmatrix} U(t)\\ V(t) \end{bmatrix} = 2 \, \mathbb{H}_\text{eff} \begin{bmatrix} U(t) \\ V(t) \end{bmatrix} , 
  \label{Eq:Ham_nh}
\end{equation}
with $\mathbb{H}_\text{eff}$ the non-Hermitian Bogoliubov De-Gennes Hamiltonian (defined in analogy with the Hermitian one).
The exponential operator $\hat{m}_j = \nep^{x \hat{n}_j}$ can be applied to a Gaussian state by simply 
applying to $\mathbb{U}$ the matrix $\mathbb{M}$, which is defined as follows:
\begin{equation}
  \begin{cases}
    \mathbb{M}_{i, i} = 1$ \quad \text{for} \ \  $i \ne j, j + L,\\
    \mathbb{M}_{j,\,j}=\mathbb{M}_{j+L,\,j+L}^{-1} = \nep^{-x}, \\
  \end{cases}
\end{equation}
    
\section{Entanglement and R\'enyi entropy in fermionic Gaussian states}
\label{subsec:entanglement}

The entanglement entropy of a subsystem of dimension $\ell$ is defined as
\begin{equation}
  S_\ell = - \rho_\ell \log \rho_\ell, \quad \mbox{with} \;\;\; \rho_\ell (t) = \text{Tr}_{L/\ell} \left[ \ket{\psi_t}\bra{\psi_t} \right]
\end{equation}
being the reduced density matrix of the subsystem. Finding $\rho_\ell (t)$ is usually a computationally hard task,
in particular for spin systems whose Hilbert space grows exponentially with the system size $L$.
Luckily, when dealing with Gaussian states (such as in the case of the Ising chain) the possibility of exploiting
Wick's theorem remarkably reduces the computational effort. In fact, it is possible to write $\rho_\ell$
by defining $\ell$ appropriate uncorrelated fermionic operators~\cite{Vidal2003,Vidal2003b}.
Below we provide details on the procedure to follow to write down these operators.

First, we need the $2L \times 2L$ correlation matrix
\begin{equation}
	\begin{aligned}
		\mathbb{G}(t)= \mathbb{U}(t)\left(\begin{array}{c|c}\mathbb{I} & 0 \\ \hline 0& 0 \end{array} \right)  \mathbb{U}^\dagger(t)= \left(\begin{array}{c|c} G(t)& F(t) \\ \hline F^\dagger(t)& 1 - G^T(t) \end{array} \right),  
	\end{aligned}
\end{equation}
with $G_{jj'}(t) \equiv \braket{\hat c_j \hat c^\dagger_{j'}}_t$
and $F_{jj'}(t) \equiv \braket{\hat c_j \hat c_{j'}}_t$. 
We now introduce the Majorana fermions
\begin{equation}
  \check{c}_{j,1} = \hat c_j^\dagger + \hat c_j, \qquad
	\check{c}_{j,2} = i ( \hat c_{j}^\dagger - \hat c_j).
\end{equation}
Analogously to the Nambu spinors, the Majorana column vector is defined as
$\check{\boldsymbol{c}} = \left( \check{c}_{1,1}, \dots, \check{c}_{L,1}, \check{c}_{1,2}, \dots, \check{c}_{L,2} \right)^T$,
through the relation
\begin{equation}
  \check{\boldsymbol{c}} = \mathbb{W} \hat \Psi, \quad \mbox{with } \;\; \mathbb{W} = \left(\begin{array}{c|c} \ \ \  \mathbb{I}& \ \mathbb{I} \\ \hline -i \mathbb{I}& i\mathbb{I} \end{array} \right).
\end{equation}

Using this relation, we can evaluate the Majorana correlation matrix
$\mathbb{M}_{nn'}(t)= \braket{\check{\boldsymbol{c}}_n\check{\boldsymbol{c}}_{n'}}$ as:
\begin{equation}
  \mathbb{M}(t) = \mathbb{W} \mathbb{G}(t) \mathbb{W}^\dagger.
\end{equation}
We can decompose the matrix $\mathbb{M}(t) = \mathbb{I} + i \mathbb{A}(t)$.
The reduced Majorana correlation matrix $\mathbb{M}^\ell$ can be then constructed according to
\begin{equation}
  \begin{cases} 
    \mathbb{M}^\ell_{n, n'} = \delta_{n, n'} + i \mathbb{A}_{n, n'}, \\
    \mathbb{M}^\ell_{n, l+ n'} = i \mathbb{A}_{n, L + n'} \\
    \mathbb{M}^\ell_{l + n, n'} = i \mathbb{A}_{L + n, n'}, \\
    \mathbb{M}^\ell_{l + n, l + n'} = \delta_{n, n'} + i \mathbb{A}_{L + n, L + n'},
  \end{cases}
\end{equation}
with $n, n' \in \{ 1, \dots , \ell\}$.

At each time step $t$, one can transform the matrix $\mathbb{A}^\ell(t)$ to a canonical form,
by a (real) orthogonal transformation $\mathbb{R}$ (Schur's decomposition)
\begin{equation}
  \mathbb{A}_\ell(t) = \mathbb{R}(t) \tilde{\mathbb{A}}(t) \mathbb{R}^\dagger(t)\,, 
  \ \text{with }\ 
  \tilde{\mathbb{A}} = \bigoplus_{q=1}^\ell \begin{pmatrix} 0 &\lambda_q \\ -\lambda_q & 0\end{pmatrix}\,.
\end{equation}
This rotation defines a new (non local) combination of Majorana fermions
$\check{\boldsymbol{d}_q} = \sum_{n = 1}^{2\ell} \mathbb{R}_{nq}(t) \check{\boldsymbol{c}}$.
Transforming back and defining $\ell$ fermionic operators $\hat d_q = \mathbb{W}^{-1} \check{\boldsymbol{d}_q}$,
it can be shown that, in this basis, the reduced density matrix factorizes in $\ell$ blocks of size $2\times2$, having eigenvalues 
\begin{equation}
	P_q(t) = \frac{1 + \lambda_q(t)}{2}, \qquad 1 - P_q(t) = \frac{1 - \lambda_q(t)}{2}\,.
\end{equation}
The entanglement entropy is thus given by 
\begin{equation}
	S_\ell(t) = - \sum_{q=1}^\ell P_q(t) \log P_q(t) + [1 - P_q(t)]\log [1 - P_q(t)]\,.
\end{equation}

In an analogous way, we can also evaluate the R\'enyi entropy of Eq.~\eqref{renna:eqn} as
\begin{equation}
  H_\beta(t) = \frac{1}{1-\beta}\sum_{q=1}^\ell \ln\left[P_q^\beta(t)+(1 - P_q(t))^\beta\right]\,.
\end{equation}

\bibliography{ising_dissipative.bib}

\end{document}